\documentclass[twocolumn,aps,floatfix]{revtex4}
\usepackage{amssymb}

\usepackage{graphicx}

\begin{document}

\title{Optical generation of quasisolitons in a single--component gas of neutral
fermionic atoms}
\author{T. Karpiuk,$^1$ M. Brewczyk,$^1$ {\L}. Dobrek$^2$, M.A. Baranov$^2$,
        M. Lewenstein$^2$, and K. Rz\c a\.zewski$^3$}                          

\affiliation{\mbox{$^1$ Uniwersytet w Bia{\l}ymstoku, ul. Lipowa 41,
                        15-424 Bia{\l}ystok, Poland}  \\
\mbox{$^2$ Institut f\"ur Theoretische Physic, Universit\"at Hannover,
           30167 Hannover, Germany} \\
\mbox{$^3$ Centrum Fizyki Teoretycznej PAN and College of Science,
           Al. Lotnik\'ow 32/46, 02-668 Warsaw, Poland}  }             

\date{\today}

\begin{abstract}
We analyze the generation of soliton--like solutions in a single--component
Fermi gas of neutral atoms at zero and finite temperatures with the phase
imprinting method. By using both the numerical and analytical calculations,
we find the conditions when the quasisolitons, which apparently resamble 
the properties of solitons in non--linear integrable equations, do exist in 
a non--interacting Fermi gas. We present the results for both spatially 
homogeneous and trapped cases, and emphasize the importance of the Fermi 
statistics and the absence of the interaction for the existence of such 
solutions.

PACS number(s): 05.30.Fk, 03.75.Fi \newline
\end{abstract}

\maketitle

\section{Introduction}

Recent pioneering work by DeMarco and Jin \cite{Jin} on realization of
quantum degeneracy in a trapped gas of fermionic $^{40}$K atoms has triggered
the interest in properties of cold fermions. As opposed to bosons, fermions
cannot occupy the same quantum state, i.e., they obey the Pauli exclusion 
principle. Since at low temperatures only the $s$--wave scattering gets 
important, the spin--polarized fermions stop to collide at lower 
temperatures and rethermalization process necessary for evaporation
breaks down. This difficulty has been overcome by trapping and cooling the
fermions in two hyperfine states, enabling in this way the evaporation by
collisions between atoms in a different spin states \cite{Jin,Jin1}. Other
groups have achieved quantum degeneracy in a gas of $^6$Li atoms by
sympathetic cooling of a mixture of bosonic ($^7$Li) and fermionic ($^6$Li)
atoms reaching the temperatures down to $0.2T_F$ \cite{Hulet,Salomon}.

Despite of experimental difficulties in achieving the strong quantum
degeneracy in fermionic gases, some theoretical work on degenerate Fermi
gases in the normal phase at zero and finite temperatures has already been
done. The static properties of one-- and two--component system were analysed
in Refs. \cite{Rokhsar}, \cite{Bruun1} and \cite{Roth}. Collective
excitations in degenerate Fermi gases in the normal phase were discussed
within the hydrodynamic approximation in Ref. \cite{Bruun} and on the basis
of the sum rules in Ref. \cite{Vichi}. In the absence of $s$--wave collisions
other forces, like for instance dipole--dipole interaction, start to play
role. The fermionic dipoles have been investigated recently, including the
analysis of stability conditions \cite{Krzysiek}. Another aspect of dynamic
behavior of a degenerate Fermi gas is studied in Ref. \cite{Tomek} where 
the question about the possibility of generation of solitons and vortices 
in a normal state of non--interacting Fermi gas is asked. It is shown by 
solving the many--body Schr\"{o}dinger equation that the phase imprinting 
method is capable of generating the solitons and vortices in such a system.
An alternative method of generating such structures is discussed in
Ref. \cite{Damski}.

So far, the phase imprinting method has been successfully applied to
generate solitons in Bose--Einstein condensates \cite{Hannover,NIST}.
Although not realized experimentally yet, it offers also a rather rare
opportunity of control of the generation of vortices \cite{Dobrek,Andrelczyk}.
The method consists in passing an off--resonant laser pulse through the 
appropriately designed absorption plate followed by the impinging it on the 
atomic gas. The laser pulse is short and strong enough so that the atomic 
motion is negligible during the pulse and the basic effect of light is a 
generation of strong optical potential $V_{opt}(\vec{r},t)$ due to the
ac--Stark  effect. This potential acts for a short time $\tau_{dur}$, and
essentially  imprints a desired phase of atomic wave functions 
$\varphi(\vec{r})=\int_0^{\tau_{dur}} V_{opt}(\vec{r},t) dt /\hbar$. 
After this, the density and the phase try to adopt to each other, which 
results in generation of structures like solitons and vortices, depending 
on the pattern written in the absorption plate.

In this paper we explore further the possibility of generating the
quasisolitons in a gas of non--interacting Fermi atoms. According to the
common knowledge regarding solitons, the nonlinearity of the problem plays
essential role, since it cancels the dispersion and allows for a propagation
of shape preserving pulses. So, at the first glance it is surprising that
solitons can exist in a system of non--interacting particles described by the
many--body Schr\"{o}dinger equation, i.e., the linear equation. However, we
show that, due to the Fermi statistics, there exists the regime of
parameters where soliton--like solutions can exist in the non--interacting
Fermi gas.

The paper is organized as follows. In Sec. \ref{Schrodinger} we solve the
many--body Schr\"odinger equation for a system of non--interacting fermions
undergoing the phase imprinting, leading to the generation of soliton--like
structures. Both zero and finite temperatures are considered. Sec. \ref
{Wigner} offers an analytical approach to the problem of propagation and
decay of solitons, based on the Wigner function formalism, for a degenerate
as well as a Boltzmann gas. In Sec. \ref{DenMatrix} another approach based
on analysis of density matrices, is followed in order to discuss the Thomas--Fermi 
approximation and to understand the role of dimensionality. Finally, we
conclude in Sec. \ref{conclusions}.

\section{Solution of the many--body Schr\"odinger equation}

\label{Schrodinger}

\subsection{Zero temperature}

At zero temperature the many--body wave function of a system of
non--interacting particles (fermions) is given by the Slater determinant with
the lowest available one--particle orbitals occupied. 
\begin{eqnarray}
\Psi (\vec{r}_1,...,\vec{r}_N) = \frac{1}{\sqrt{N!}} \left | 
\begin{array}{lllll}
\varphi_1(\vec{r}_1) & . & . & . & \varphi_1(\vec{r}_N) \\ 
\phantom{aa}. &  &  &  & \phantom{aa}. \\ 
\phantom{aa}. &  &  &  & \phantom{aa}. \\ 
\phantom{aa}. &  &  &  & \phantom{aa}. \\ 
\varphi_N(\vec{r}_1) & . & . & . & \varphi_N(\vec{r}_N)
\end{array}
\right | \, . \label{Slater}
\end{eqnarray}
The one--particle orbitals $(\varphi_1(\vec{r}),...,\varphi_N(\vec{r}))$ are
orthogonal and they undergo unitary evolution. Hence, their orthogonality is
sustained, both during the phase imprinting and afterwords. Therefore, the
one--particle density matrix is at any time given by the following formula 
\begin{equation}
\rho_1(\vec{r}^{\, \prime}, \vec{r}^{\, \prime \prime}, t) = \frac{1}{N}
\sum_{i=1}^N \varphi_i(\vec{r}^{\, \prime}, t) \; \varphi_i^*(\vec{r}^{\,
\prime \prime}, t) \, ,
\end{equation}
and its diagonal part is just the particle density.

Evolution of each single--particle orbital is obtained from the
one--particle Schr\"odinger equation and can be split into two stages.
First, the phase imprinting technique is used to disturb the system in a way
of writing a desired phase on it. Assuming fast enough phase imprinting,
the resulting wave function has the form (this procedure can be also performed 
numerically within the finite duration of the imprinting process): 
\begin{eqnarray*}
\varphi_{k}(\vec{r}) \longrightarrow \varphi^{\prime}_{k}(\vec{r}) =
\varphi_{k}(\vec{r}) \, \exp(i \phi(x)) \, ,
\end{eqnarray*}
for $k=1,...,N$, where $\phi(x)$ is the phase imprinted on each atom.
Throughout this paper we deal with the generation of quasisolitons, we assume
that the imprinted phase depends only on a single coordinate, say '$x$', and 
utilize $\phi(x) \sim(1+\tanh(x/d_{\mathrm{phase}}))$. We consider a Fermi gas 
in a three--dimensional box or harmonic trap characterized by the frequencies 
$\omega_x$, $\omega_y$, and $\omega_z$. Both are separable in Cartesian 
coordinates. The one--particle orbitals $\varphi_{k}(\vec{r})$ are taken as 
a product $\, \varphi_{n_x}^{(1)} (x) \,\varphi_{n_y}^{(2)} (y) \, 
\varphi_{n_z}^{(3)} (z)$ of eigenvectors of the Hamiltonians of 
one--dimensional boxes or harmonic oscillators.

In the second stage the system undergoes the free evolution in
three--dimensional space. However, because the Hamiltonian separates in
coordinates '$x$', '$y$', and '$z$', the time propagation of each orbital 
can be easily reduced to one--dimensional case: 
\begin{eqnarray*}
\varphi_{k}(x,y,z,t) &=& e^{- i \mathrm{{E}_{n_y} t /\hbar}} \: e^{- i 
\mathrm{{E}_{n_z} t /\hbar}} \\
&\times& \varphi_{n_y}^{(2)} (y,0) \, \varphi_{n_z}^{(3)} (z,0) \\
&\times& \varphi_{n_x}^{(1)} (x,t)
\end{eqnarray*}
where 
\begin{eqnarray*}
\varphi_{n_x}^{(1)} (x,t) &=& e^{- i \mathrm{{H_x} \,t /\hbar}}
\varphi_{n_z}^{(1)} (x,0) \, e^{i \phi(x)}
\end{eqnarray*}
\noindent and $\mathrm{H_x}$ is the Hamiltonian of one--dimensional box or
harmonic oscillator. The diagonal part of one--particle density matrix is
given then by the expression: 
\begin{eqnarray}
\rho(\vec{r}, t) = \frac{1}{\mathrm{N}} \, \sum_{n_x, n_y, n_z} \,
|\varphi_{n_y}^{(2)} (y,0)| ^2 \; |\varphi_{n_z}^{(3)} (z,0)| ^2  \nonumber
\\
\nonumber \\
\times \, |\varphi_{n_x}^{(1)} (x,t)| ^2 \,. \phantom{aaaaaaaaa}
\label{3Dden}
\end{eqnarray}

\begin{figure}[tbp]
\resizebox{2.9in}{2.4in}
{\includegraphics{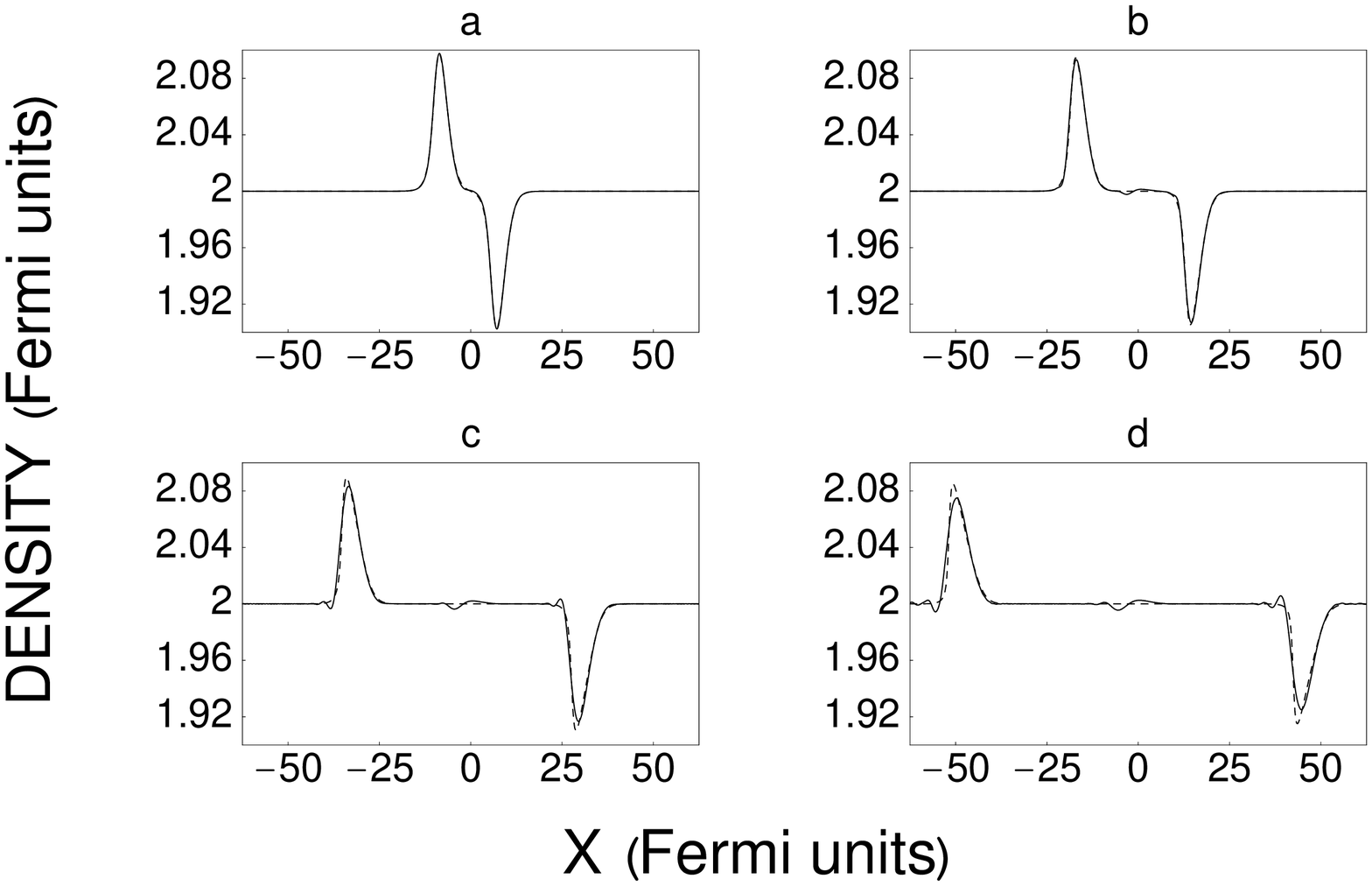}}
\caption{Images of the fermionic density (normalized to the number of atoms)
of N=$500$ atoms after writing a single phase step of $1.0\,\pi$ and the
width $d_{\mathrm{phase}}=2.5\, \lambda_F$ onto a one--dimensional
non--interacting Fermi gas confined in a box at different times: (a) 24.7, (b)
49.3, (c) 98.7, and (d) 148.0 in Fermi units ($\hbar/\varepsilon_F$). Each
graph consists of two curves: one obtained based on the many--body
Schr\"odinger equation (solid line) and the second coming from the
one--dimensional Thomas--Fermi model (dashed line).}
\label{1Dbox}
\end{figure}

In Fig. \ref{1Dbox} we plotted the density profiles of a uniform system of $%
500$ atoms after imprinting the phase designed in the form $\phi(x)=\phi_0
(1+\tanh(x/d_{\mathrm{phase}}))/2$. The characteristic length (the Fermi
length) is given by the formula $\lambda_F=h/p_F=2/N L$, where $p_F$ is the
Fermi momentum and $L$ is the size of one--dimensional box. For the case of
Fig. \ref{1Dbox}, the Fermi length equals $0.004\,L$ and is $2.5$ times
smaller than the width characterising the jump of the imprinted phase. As
discussed more precisely in Sec. \ref {Wigner}, the evolution of the density
after phase imprinting depends strongly on the ratio
$d_{\mathrm{phase}}/\lambda_F$ between the width of the phase step and the 
Fermi length.  

\begin{figure}[tbp]
\resizebox{2.9in}{2.4in}
{\includegraphics{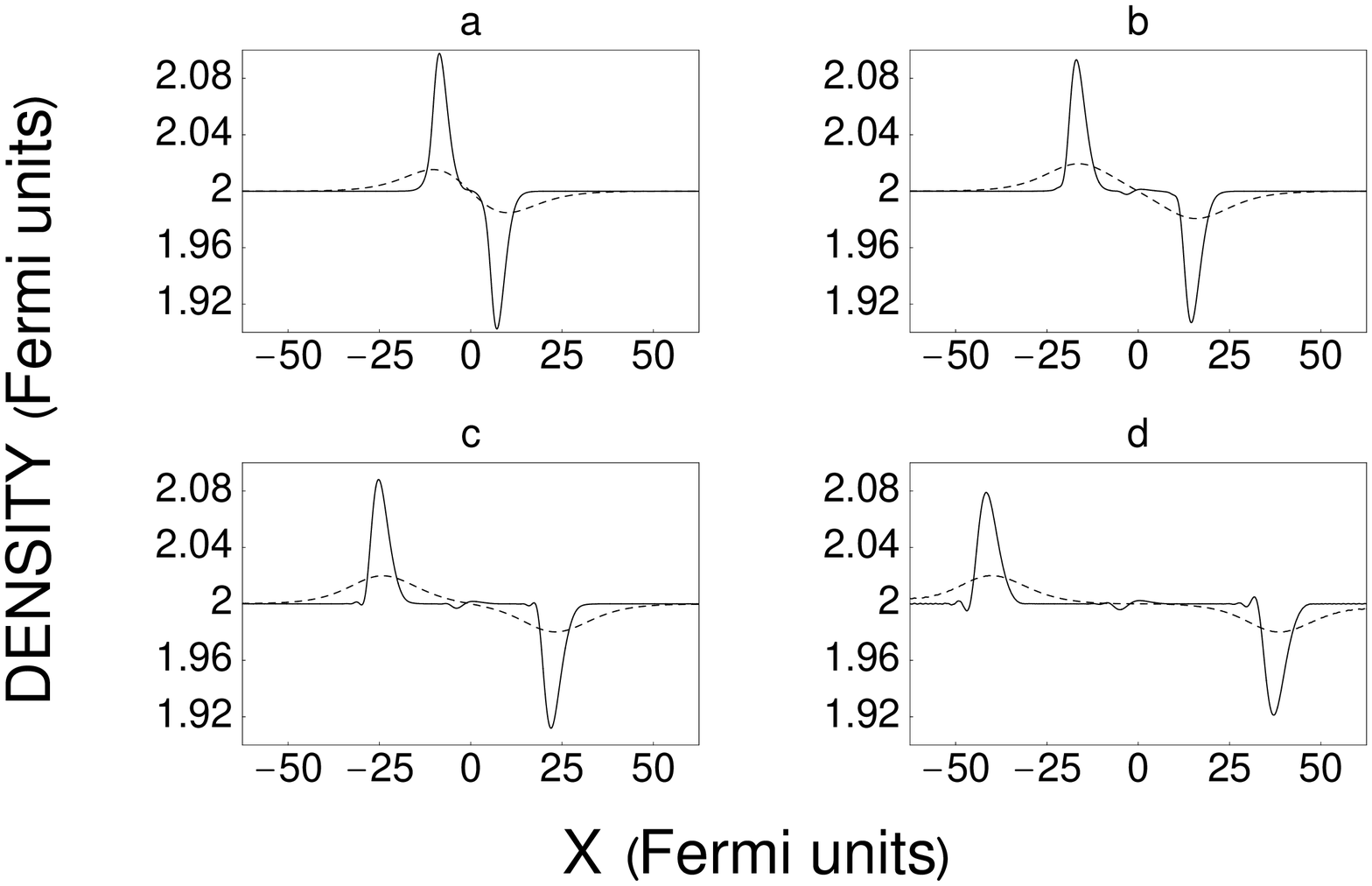}}
\caption{Comparison of bright--dark soliton structures for various widths of
imprinted phase: 2.5 $\lambda_F$ (solid line) and 12.5 $\lambda_F$ (dashed
line). The snapshots are taken at times (a) 24.7, (b) 49.3, (c) 74.0, and
(d) 123.3 in Fermi units. Other parameters are the same as in Fig. \ref
{1Dbox}.}
\label{1Dapprox}
\end{figure}

It is clear from Fig. \ref{1Dbox} (as well as Fig. \ref{1Dapprox}) that two
quasisolitons (the bright and the dark one) are generated. They propagate in
opposite directions with distinct velocities. We present in Table \ref
{table1} the values of soliton's velocities for various widths of the phase
steps. The speed of the bright quasisoliton is always higher than the speed of
sound whereas dark quasisolitons move slower than the sound. However,
increasing the ratio $d_{\mathrm{phase}}/\lambda_F$ both velocities are
getting closer to the speed of sound ($p_F/m$ in one--dimensional space),
reaching eventually (i.e., in the limit $d_{\mathrm{phase}}\gg\lambda_F$) the
value $p_F/m$ as shown in Secs. \ref{Wigner} and \ref{DenMatrix}. In Sec.
\ref{DenMatrix} we also analyze the $d_{\mathrm{phase}}\gg\lambda_F$ limit by
decreasing $\lambda_F$ (increasing the number of atoms) and solving the
equations of the Thomas--Fermi model.

\begin{table}[tbp]
\caption{Speeds of dark and bright quasisolitons (in units of speed of sound)
generated in a one--dimensional homogeneous Fermi gas of $500$ atoms after
imprinting a single phase step of $\pi$ and various widths $d_{\mathrm{phase}}$
(in units of the Fermi length).}
\label{table1}%
\begin{ruledtabular}
\begin{tabular}{ccc}
$\frac{d_{\mathrm{phase}}}{\lambda_F}$ & $\frac{v_b}{c}$ & $\frac{v_d}{c}$ \\
\hline
\hspace{0.13in}1.25 & -1.061 & +0.953 \\
\hspace{0.13in}2.50  & -1.044 & +0.958 \\
\hspace{0.13in}5.00  & -1.033 & +0.963 \\
12.5  & -1.014 & +0.981 \\
\end{tabular}
\end{ruledtabular}
\end{table}

\begin{figure}[tbp]
\resizebox{2.9in}{2.2in}
{\includegraphics{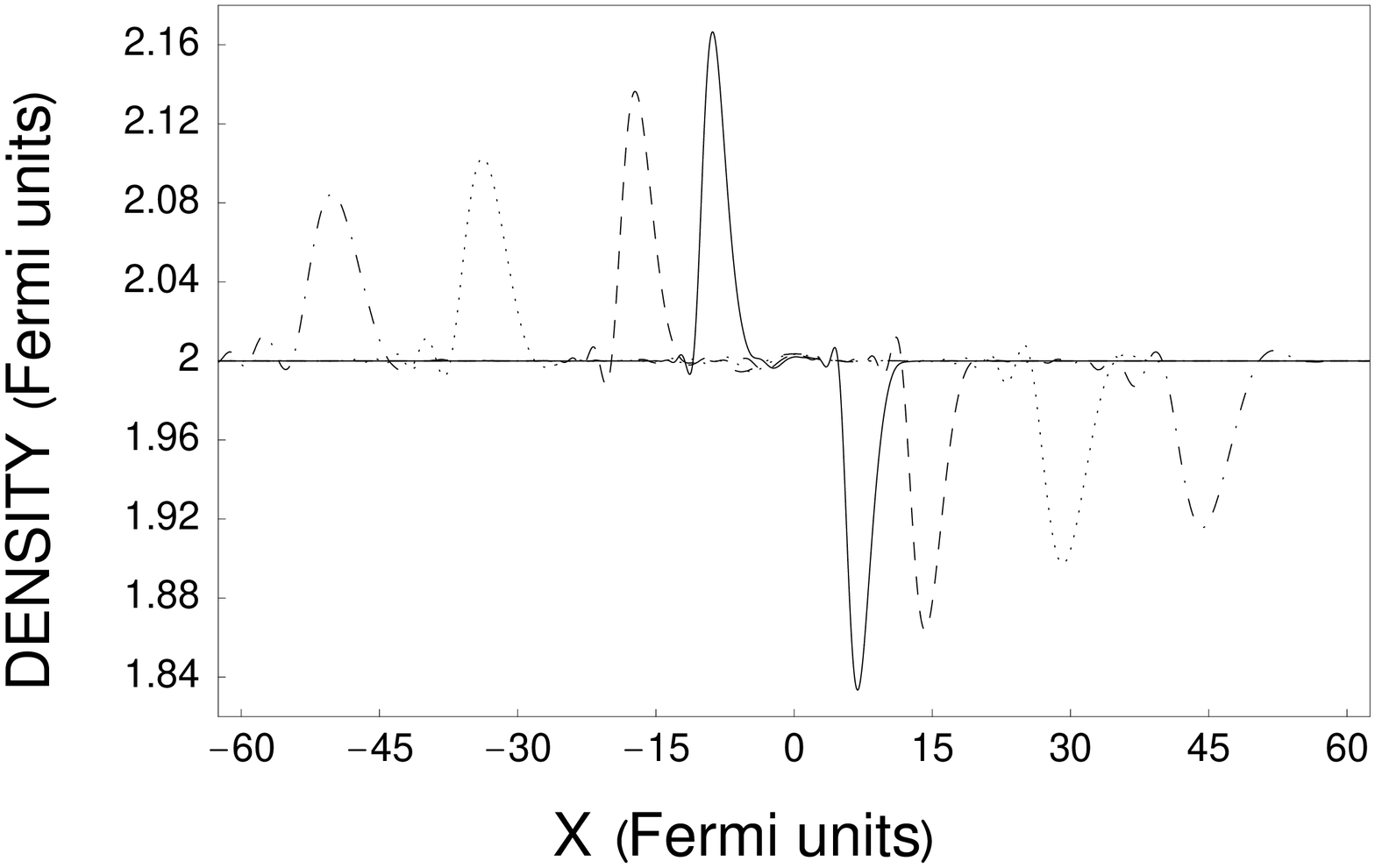}}
\caption{Soliton--like structures in the regime where $d_{\mathrm{phase}%
}\lesssim \lambda _F$ (here, $d_{\mathrm{phase}} = 1.25\, \lambda _F$). The
snapshots are taken at times (a) 24.7 (solid line), (b) 49.3 (dashed line),
(c) 98.7 (dotted line), and (d) 148.0 (dashed--dotted line) in Fermi units.
Other parameters are the same as in Fig. \ref{1Dbox}.}
\label{1Ddecay}
\end{figure}

Fig. \ref{1Dapprox} shows that going deeper into the regime $%
d_{\mathrm{phase}}\gg\lambda_F$ the soliton structures become less pronounced.
This again is explained in terms of Wigner function in Sec. \ref{Wigner} and
based on Thomas--Fermi approach in Sec. \ref{DenMatrix}. The condition 
$d_{\mathrm{phase}}\gg\lambda_F$ means that the height (or depth) of solitons 
is getting smaller in comparison with unperturbed density. Going in opposite 
direction (i.e., $d_{\mathrm{phase}}\lesssim \lambda_F$) results, as presented 
in Fig. \ref{1Ddecay}, in a change of shape of the structures (dispersion)
on a time scale of the order of $\hbar/\varepsilon_F$. 

Another aspect of Fig. \ref{1Dbox} is connected with the Thomas--Fermi
approximation discussed in Sec. \ref{DenMatrix}. Each frame of Fig. \ref
{1Dbox} shows, in fact, two curves; one of them (solid line) is obtained by
solving the many--body Schr\"odinger equation, and the second (dashed line) is
representing the Thomas--Fermi approach. Both curves match well which proves
the validity of the Thomas--Fermi approximation in one--dimensional space
for large enough number of atoms just as it happens in the case of a
harmonic trap (see Fig. 1 in Ref. \cite{Tomek}, where two-- and
three--dimensional cases are also discussed).

\subsection{Finite temperatures}

One way of including temperature effects on the soliton's generation and
propagation is to allow for populating the single--particle states of higher
energy than the Fermi energy. We have realized this by considering a grand
canonical ensemble for a system of non--interacting fermions. We generate a
number of many--particle configurations according to the Fermi--Dirac
statistics, i.e., with the probability of populating a particular
one--particle state $\varphi_k$ given by $\exp(\beta (\varepsilon_k - \mu))
+ 1)^{-1}$, where $\varepsilon_k$ is the state energy, $\beta$ determines
the bath temperature (via $\beta = 1/ k_B T$), and $\mu$ is the chemical
potential (dependent on temperature in general). More precisely, for each
one--particle state (unless the probability of populating this state is
lower than $10^{-6}$) we generate a random number from the interval $[0,1]$.
If the number is less than $\exp(\beta (\varepsilon_k - \mu)) + 1)^{-1}$,
the one--particle state is populated, otherwise it is left empty. Here, 
$\varepsilon_k$ is the energy of the state under consideration, the
chemical potential 
\begin{eqnarray}
\mu=\varepsilon_F[1-(\pi^2/3) (k_B T/\varepsilon_F)^2] \, , & & 
\label{3Dchempot}
\end{eqnarray}
and $\varepsilon_F$ equals the Fermi energy corresponding to the average
number  of atoms. The above formula for the chemical potential applies for 
the Fermi gas trapped in a harmonic potential for the temperatures up to 
$0.55\, T_F$ \cite{Rokhsar}.

In such a way, having given the temperature and the average number of atoms,
the many--particle configurations are sampled. Then, for each
many--particle configuration the phase imprinting is performed followed by a
free evolution. Finally, the density is calculated as an average according
to a grand canonical ensemble rules. We have investigated in this way the
propagation of quasisolitons for temperatures up to $0.3\, T_F$, already
attainable in experiment \cite{Jin1,Hulet,Salomon}.

\begin{figure}[tbp]
\resizebox{2.9in}{2.4in}
{\includegraphics{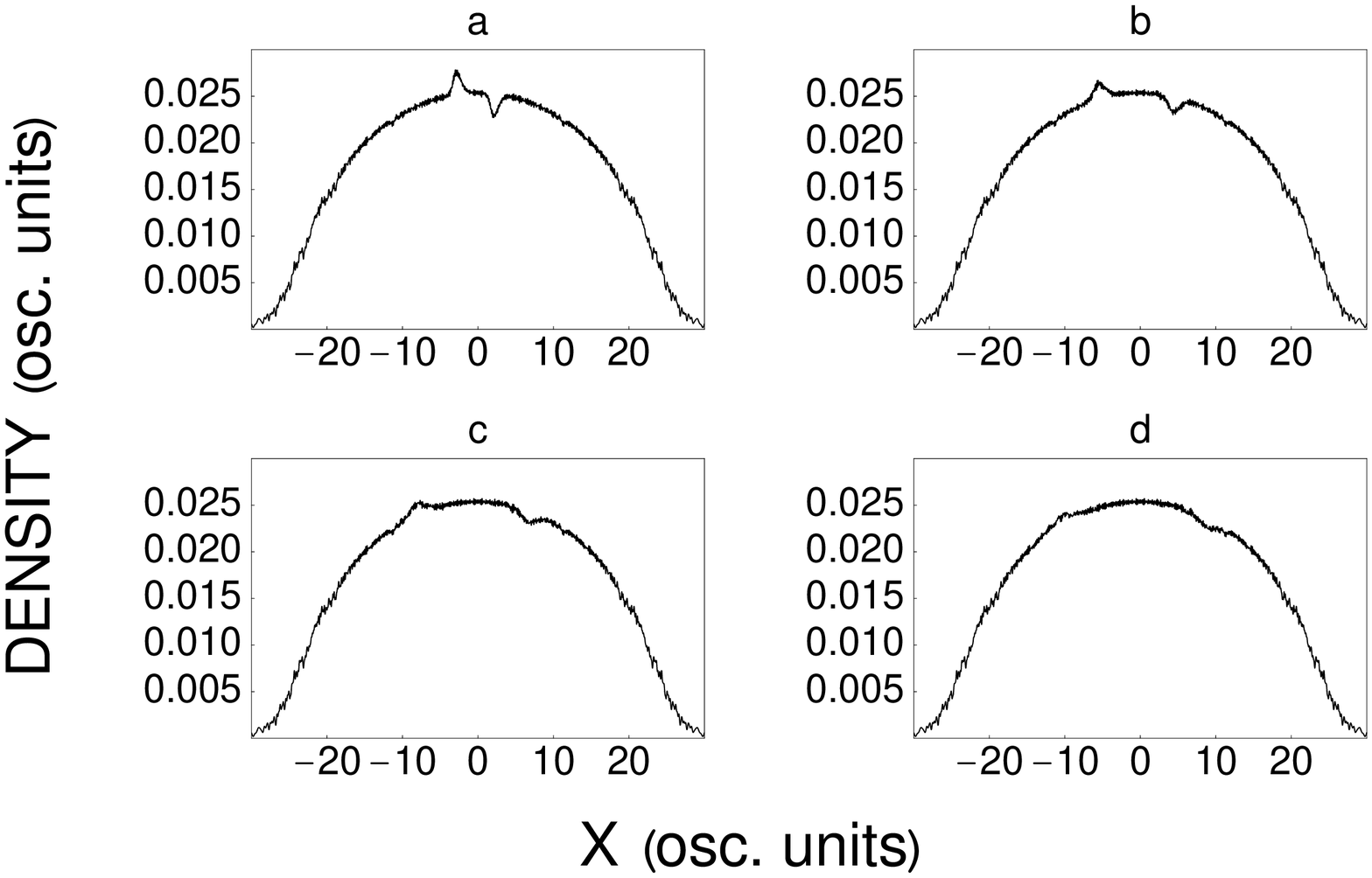}}
\caption{Evolution of the density distribution of 300 (average number) Fermi
atoms in one--dimensional harmonic trap at temperature $0.2\, T_F$ after
imprinting a single phase step of $2.0\,\pi$ and $d_{\mathrm{phase}}=0.5$ osc.
units. The successive frames correspond to moments: (a) 0.1, (b) 0.2, (c)
0.3, and (d) 0.4 in units of $1/\omega$.}
\label{1Dosctemp1}
\end{figure}

\begin{figure}[tbp]
\resizebox{2in}{3in}
{\includegraphics{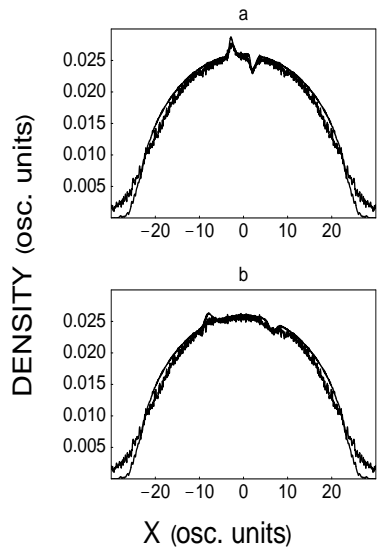}}
\caption{Comparison of bright--dark soliton structures at different
temperatures: $0.1\, T_F$ (less rugged curve, averaged over $1000$
configurations) and $0.3\, T_F$ (more rugged curve, averaged over 
$10000$ configurations). Frames (a) and (b) correspond to moments 
$0.1$ and $0.3$ in units of $1/\omega$ respectively and other 
parameters are the same as in Fig. \ref{1Dosctemp1}.}
\label{1Dosctemp2}
\end{figure}

\begin{figure}[tbp]
\resizebox{2.9in}{2.3in}
{\includegraphics{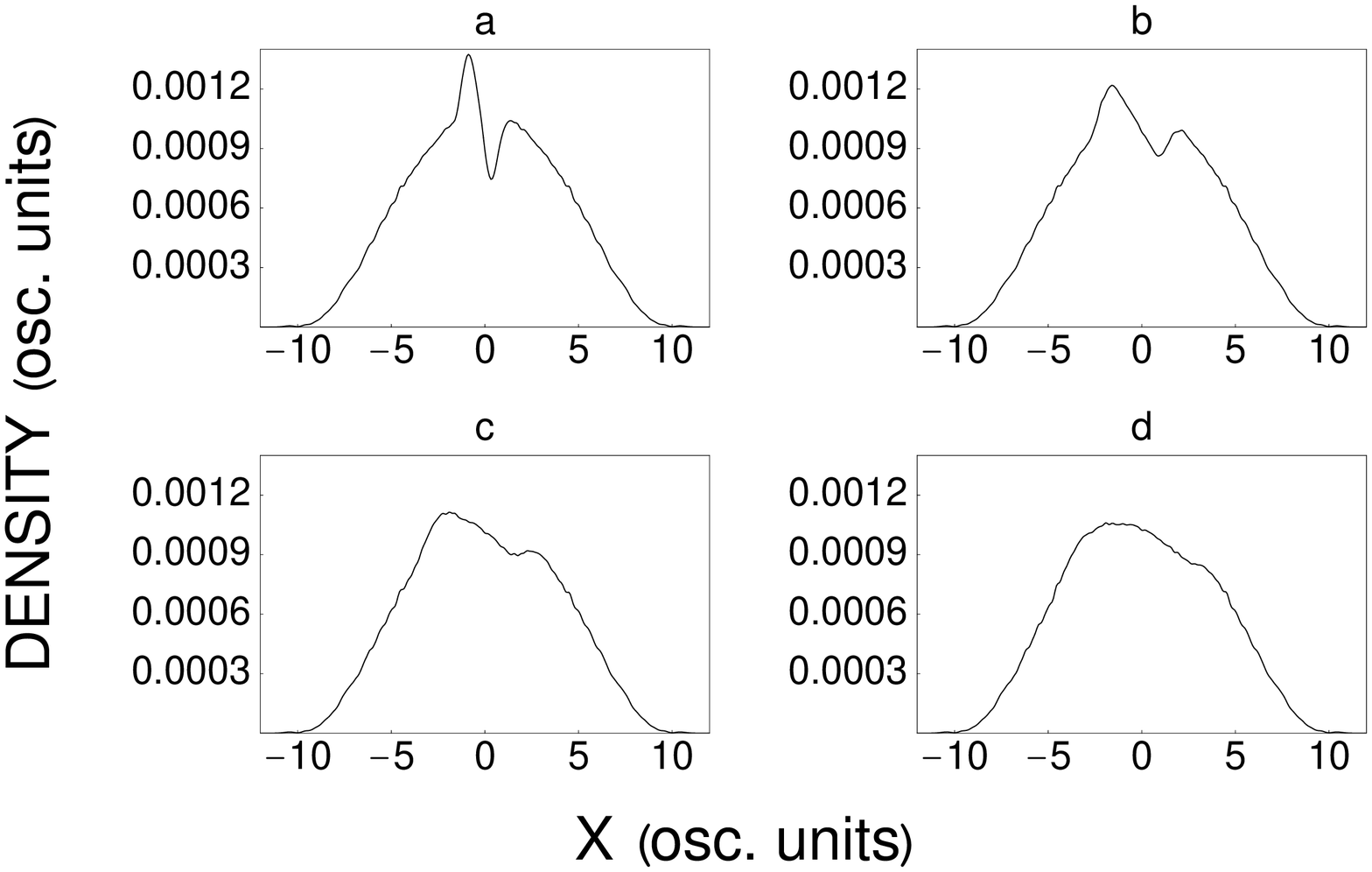}}
\caption{Density profiles of non--interacting fermionic gas confined in
three--dimensional spherically symmetric harmonic trap at different times:
(a) 0.1, (b) 0.2, (c) 0.3, and (d) 0.4 in units of $1/\omega$ after writing
a single phase step of $\phi_0=1.0\,\pi$ and $\zeta=0.2$ \mbox{osc. units}.
The average number of atoms N=$10^4$, the temperature is equal to $0.1\, T_F$%
, and $1000$ configurations were used in the averaging procedure.}
\label{1Dosctemp3}
\end{figure}

In Fig. \ref{1Dosctemp1} we plot evolution of the averaged density of the
Fermi gas confined in a one--dimensional harmonic trap of frequency
$\omega$ after phase imprinting of a single step of $2 \pi$ and the width
equal to $0.5$ \mbox{osc. units.} ($\sqrt{\hbar/(m\omega)}$). Here, the
chemical potential is given by the formula $\mu=\varepsilon_F+k_B T
\log[1-\exp(-\varepsilon_F/k_B T)]$, which is a one--dimensional
counterpart of formula (\ref{3Dchempot}), the temperature equals $0.2\, T_F$,
and the averaging procedure is performed over $5000$ many--body configurations.
Both, bright and dark solitons are still present in the system exhibiting
basically the same properties as at zero temperature. For higher temperatures
(already $0.3\, T_F$) the soliton structures start to get hidden within the
noise originating in the averaging procedure (Fig. \ref{1Dosctemp2}).
However, one has to remember that rather a particular many--body
configuration is realized in a single experiment, and not an ensemble of them.
Finally, Fig. \ref{1Dosctemp3} gives an example that going from one-- to
three-dimensional space, predictions made based on lower dimensionality
survive. Fig. \ref{1Dosctemp3} shows density profiles of fermionic gas
trapped in a spherically symmetric potential (averaged over $1000$
configurations at $0.1\, T_F$ temperature) and can be compared with
corresponding figure from Ref. \cite{Tomek} at zero temperature since
the details of phase imprinting are the same. 

\section{Wigner function approach}

\label{Wigner}

The Wigner function $w(x,p,t)$ is defined as

\begin{equation}
w(x,p,t)=\int dy\left\langle \hat{\psi} ^{\dagger }(x+y/2,t) \hat{\psi}
(x-y/2,t)\right\rangle \exp (-ipy) \;,  \label{def}
\end{equation}
where $\hat{\psi}^{\dagger }(x,t)$ and $\hat{\psi}(x,t)$ denote atomic
creation and annihilation operators in the Heisenberg picture, respectively.
It obeys the equation

\begin{equation}
\left( \partial _{t}+\frac{p}{m}\partial _{x}-\partial _{x}V_{ext}(x)\partial
_{p}\right) w(x,p,t)=0 \;,  \label{eqmotion}
\end{equation}
where $V_{ext}(x)$ is the external harmonic trap potential. It immediately 
follows from the
definition, Eq. (\ref{def}), that the density of particles at time $t$ is

\begin{equation}
\rho(x,t)=\int \frac{dp}{2\pi \hbar } \, w(x,p,t) \;.  \label{density}
\end{equation}

The phase imprinting at $t=0$ modifies the Wigner function in the following
way: 
\begin{eqnarray}  \label{phaseimp}
w_{0}(x,p,t=0)\rightarrow \phantom{aaaaaaaaaaaaaaaaaaaaaaaaaaaaaai}
\nonumber \\
w_{0}(x,p)=\int dy\left\langle \hat{\psi} ^{\dagger }(x+y/2,t=0) \hat{\psi}
(x-y/2,t=0)\right\rangle  \nonumber \\
\times \exp (-ipy/\hbar - i\phi (x+y/2)+i\phi (x-y/2)) \;,  \nonumber
\\
\end{eqnarray}
where $\phi (x)$ is the imprinted phase. The function $w_{0}(x,p)$
provides the initial condition for the dynamical evolution according to Eq. (%
\ref{eqmotion}).

Let us consider first the spatially homogeneous case ($V_{ext}(x)=0$). Eq. (\ref
{eqmotion}) with the initial condition $w(x,p,t=0)=w_{0}(x,p)$ can be easily
solved. The solution is

\begin{equation}
w(x,p,t)=w_{0}(x-(p/m)t,p) \;,  \label{solution}
\end{equation}
and reflects the ballistic motion of particles in the absence of an external
potential. By using the fact that in the spatially homogeneous case the
correlation function $\left\langle \hat{\psi} ^{\dagger }(x+y/2,t=0) \,
\hat{\psi} (x-y/2,t=0)\right\rangle =\rho_{0}(y)$, i.e., it depends only
on '$y$', the expression for $w_{0}(x,p)$ can be rewritten in the form 
\begin{eqnarray}
w_{0}(x,p)=\int dy\rho _{0}(y)\exp (-ipy/\hbar -i\phi (x+y/2)  \nonumber
\\
+i\phi (x-y/2)) \;. \phantom{aaaaa}  \label{solution1}
\end{eqnarray}

The above formulae have a general character, and are valid for both
degenerate and non-degenerate (Boltzmann) Fermi gas. The difference between
these two cases is only in the specific form of the function $\rho _{0}(y)$.

Usually in the experiments, the imprinted phase $\phi (x)$ is a function 
with transition regions of the size of the order of $d_{\mathrm{%
phase}}$ between the regions with constant values of the phase. Depending on
the ratio between $d_{\mathrm{phase}}$ and the characteristic width of the
function $\rho _{0}(y)$, one has completely different behavior of the
solution Eq. (\ref{solution}) and, therefore, the density evolution. In the
degenerate regime, the characteristic width of $\rho _{0}$ is given by the
Fermi wavelength $\lambda _{F}$, which is inversely proportional to the Fermi
momentum, $\lambda _{F}=h /p_{F}$. In the opposite case of a Boltzmann
gas, the width is given by the thermal wavelength $\lambda _{T}=h/\sqrt
{2mT}$.

We first analyze the case where $d_{\mathrm{phase}}\gg \lambda _{F},\
\lambda _{T}$. Under this condition, one can expand the exponent in Eq. (\ref
{solution1}) in powers of $y$, and leave only the lowest, linear in $y$,
contribution: 
\begin{eqnarray}
w_{0}(x,p) & \approx & \int dy\rho _{0}(y)\exp (-ipy/\hbar -i\phi
^{\prime}(x)y)  \nonumber \\
& = & \widetilde{\rho }_{0}(p+\phi ^{\prime }(x)) \;,  \label{solution2}
\end{eqnarray}
where $\widetilde{\rho }_{0}(p)$ is the Fourier transform of the function $%
\rho _{0}(y)$ and $\phi ^{\prime }$ is the derivative of $\phi $. The
density of the particles, Eq. (\ref{density}), can then be written as 
\begin{eqnarray}
\rho(x,t) & = & \int \frac{dp}{2\pi \hbar }w_{0}(x-(p/m)t,p)  \nonumber \\
& \approx & \int \frac{dp}{2\pi \hbar} \widetilde{\rho }_{0}(p+\hbar \phi
^{\prime }(x-(p/m)t)) \;.  \label{density1}
\end{eqnarray}
Keeping in mind that $\phi ^{\prime }\lesssim \pi /d_{\mathrm{phase}}$
and, therefore, is much smaller than the characteristic momenta, $p_{F}$ and 
$p_{T}\sim \sqrt{mT}$ for the degenerate and Boltzmann gas, respectively,
Eq. (\ref{density1}) can be simplified as: 
\begin{equation}
\rho(x,t)\approx \rho_{0}+\int \frac{dp}{2\pi }\widetilde{\rho }_{0}^{\,\prime
}(p)\phi ^{\prime }(x-(p/m)t),  \label{density2}
\end{equation}
where $\rho_{0}=\int (dp/(2\pi\hbar) )\widetilde{\rho }_{0}(p)$ is the initial
constant density and $\widetilde{\rho }_{0}^{\,\prime }$ is the derivative of $%
\widetilde{\rho }_{0}$. The function $\phi ^{\prime }(x)$ is only nonzero
in the transition regions of the width $d_{\mathrm{phase}}$. Therefore, Eq. (%
\ref{density2}) has a very clear meaning. It represents a superposition of
bumps moving with different velocities $p/m$ which are distributed with the
weights $\widetilde{\rho }_{0}^{\prime }(p)$.

For a degenerate Fermi gas at zero temperature one has $\widetilde{\rho }%
_{0}^{\,\prime }(p)=\delta (p+p_{F})-\delta (p-p_{F})$, and, hence, 
\begin{equation}
\rho(x,t)\approx \rho_{0} + \frac{1}{2\pi} [\phi
^{\prime}(x+(p_{F}/m)t)-\phi ^{\prime} (x-(p_{F}/m)t)] \;.
\label{density3}
\end{equation}
The above formula describes a ''soliton-like'' motion (without broadening and 
interaction) of bright and dark structures in the density, described by the
function $\phi ^{\prime }(x)$, in the opposite directions with the same
velocity $v_{F}=p_{F}/m$. The physical explanation of this phenomenon is
very simple. Under the condition $d_{\mathrm{phase}}\gg \lambda _{F}$, the
quantum state of the system after the phase imprinting contains excitations,
particle-hole pairs, only in the vicinity of the Fermi surface. (The energy
width of the excitation distribution is of the order of $\hbar /d_{\mathrm{%
phase}}\ll p_{F}$.) As a result, they have the same velocity. either $v_{F}$
or $-v_{F}$, and, therefore, the corresponding density profile moves in both
directions with velocities $\pm v_{F}$, respectively, without any
distortion. To be precise, the broadening in this case is of the second
order in $d_{\mathrm{phase}}/\lambda _{F}$. To determine the characteristic
time scale $\tau _{F0}$, one has to keep in mind that the width of the
distribution of the excitation momenta after the phase imprinting is of the
order of $\delta p\sim \hbar /d_{\mathrm{phase}}\ll p_{F}$. Hence, the time $%
\tau _{F0}$ can be estimated as $\tau _{F0}\sim (\hbar /T_{F})(d_{\mathrm{%
phase}}/\lambda _{F})^{2}\gg \hbar /T_{F}$.

\begin{figure}[tbp]
\resizebox{2.9in}{5.0in}
{\includegraphics{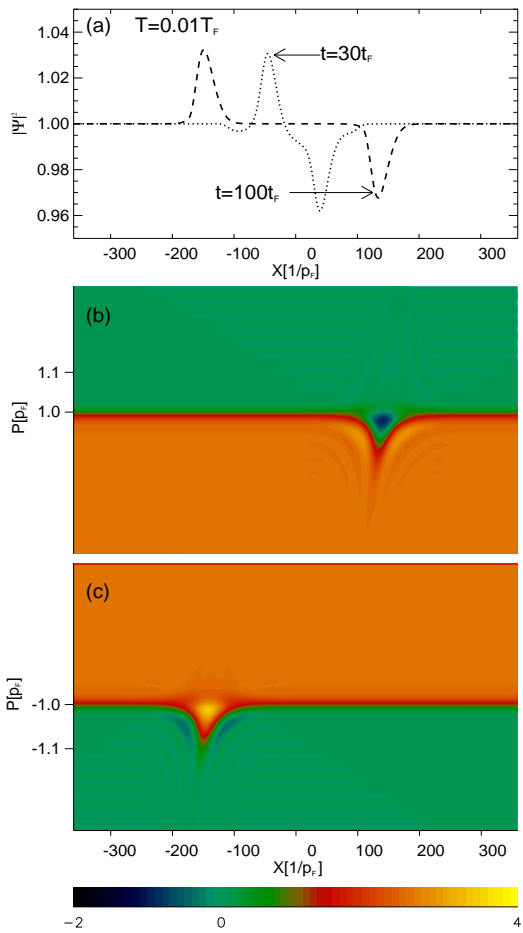}}
\caption{The time evolution of the gas density and the Wigner function after the
phase imprinting with $d_{\mathrm{phase}}=160\lambda _{F}$ at temperature 
$T=0.01T_{F}$: (a) the density profile at the times $t=30t_{F}$ (the dotted
curve) and $t=100t_{F}$ (the dashed curve), respectively, where $t_{F}=\hbar
/T_{F}$; (b) and (c) the Wigner function for the time $t=100t_{F}$, where
$t_{F}=\hbar/\varepsilon_F$.}
\label{fig1}
\end{figure}

At temperatures $T\ll T_{F}$ (but $T>T_{F}\lambda _{F}/d_{\mathrm{phase}}$,
the condition when the thermal width of the excitation distribution is
larger then that after the phase imprinting), the situation is very similar,
and Eq. (\ref{density2}) can be written in the form

\begin{eqnarray}
\rho(x,t) &\approx &\rho_{0}+\int \frac{ds}{2\pi }f_{F}(s,T)\left[ \phi
^{\prime }(x+t(p_{F}+s)/m))\right.   \nonumber \\
&&\left. -\phi ^{\prime }(x-t(p_{F}+s)/m)\right] \;,  \label{density4}
\end{eqnarray}
where $f_{F}^{\prime }(s,T)=(p_{F}/4mT)\cosh ^{-2}(sp_{F}/2mT)$ is related
to the derivative of the Fermi-Dirac distribution and describes broadening
of the $\delta $-like peaks at $p=\pm p_{F}$ in the distribution $\widetilde{
\rho }_{0}^{\,\prime }(p)$ at finite temperatures. As a result, now one has a
superposition of bright bumps and dark wells in the density profile, moving with
slightly different velocities distributed with two narrow peaks at $\pm
p_{F}/m$ with the width of the order of $mT/p_{F}\sim p_{F}T/T_{F}\ll p_{F}$.
This results in spreading of bumps and wells in the density profile on a
time scale $\tau _{FT}\sim (\hbar /T)(d_{\mathrm{phase}}/\lambda _{F})\sim
(\hbar /T_{F})(T_{F}/T)(d_{\mathrm{phase}}/\lambda _{F})\gg \hbar /T,\ \hbar
/T_{F}$ (at $T=T_{F}\lambda _{F}/d_{\mathrm{phase}}$ one has $\tau
_{FT}=\tau _{F0}$). Therefore, the time evolution of the gas density after a
phase imprinting can be described in terms of a combination of several
quasisolitons (bumps) and quasi--darksolitons (wells) which move through
each other without any interaction and, for times $t\ll \tau $, decay
(broaden) very slowly. In Fig. \ref{fig1} we present the results of numerical 
calculations of the time evolution of the density and Wigner functions for a 
spatially homogeneous gas at the temperature $T=0.01T_F$ after the 
$2\pi$--step phase imprinting with $d_{\mathrm{phase}}=160 \lambda_F$. 
The results clearly demonstrate (see Fig. \ref{fig1}a) that under these 
conditions the soliton--like structures caused by the phase imprinting 
behave like real solitons, i.e., they propagate without any noticeable
decay (broadening) on a long time scale. The blue regions in Figs. 
\ref{fig1}b and \ref{fig1}c correspond to negative values of the Wigner
function, and, therefore, indicate the quantum origin of the phenomenon.

\begin{figure}[tbp]
\resizebox{2.9in}{2.3in}
{\includegraphics{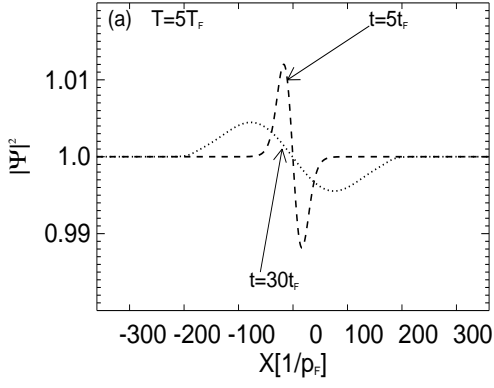}}
\caption{The density of the gas with temperature $T=5T_{F}$ at time $t=5t_{F}$
(the dashed curve) and $t=30t_{F}$ (the dotted curve), respectively, after
the phase imprinting with $d_{\mathrm{phase}}=160\lambda _{F}$. }
\label{fig2}
\end{figure}

For high temperatures $T\gg T_{F}$ gas is in the Boltzmann regime, and Eq. 
(\ref{density2}) takes the form

\begin{equation}
\rho(x,t)\approx \rho_{0}+\int \frac{dp}{2\pi }f_{B}(p,T)\phi ^{\prime
}(x-(p/m)t)\;,  \label{density5}
\end{equation}
where $f_{B}(p,t)=\rho_{0}\sqrt{2\pi /mT}(p/mT)\exp (-p^{2}/2mT)$ describes the
distribution with two broad peaks at $p=\pm p_{T}$ with $p_{T}\sim \sqrt{mT}$%
, with the width $\delta p\sim p_{T}$. In this case, the decay of the
original density profile takes place more rapidly than in the case of a
degenerate gas, on the time scale $\tau _{B}\sim (\hbar /T)(d_{\mathrm{phase}%
}/\lambda _{T})\sim \tau _{F}\sqrt{T_{F}/T}\gg \tau _{F}$. The corresponding 
numerical results for the density evolution at the temperature $T=5T_F$ are 
shown in Fig. \ref{fig2}. Although the parameters of the phase imprinting are 
exactly the  same as for the case of Fig. \ref{fig1}, the temperature (or, in 
other words, the absence of the sharp Fermi surface) cause a rapid decay of 
soliton-like structures.      

The same consideration is applied to the trapped gas in the harmonic
potential. In this case, due to the periodicity of the motion, the essential
scale for the time evolution is set by the inverse level spacing: $t\leq
2\pi /\omega $, where $\omega $ is the trap frequency. Therefore, after
phase imprinting one would observe a quasisoliton behavior if $2\pi /\omega
\ll \tau _{F0},\ \tau _{FT}$, or $2\pi /\omega \ll \tau _{B}$ for the
degenerate or Boltzmann regime, respectively. More explicitly, in the
degenerate regime one has the following requirement for the temperature $T$
and the width $d_{\mathrm{phase}}$: $[\hbar \omega /\max (T,T_{F}\lambda
_{F}/d_{\mathrm{phase}})](d_{\mathrm{phase}}/\lambda _{F})\gg 1$, or $d_{%
\mathrm{phase}}/l_{0}\gg \max (1,T/\hbar \omega \sqrt{N})$, where $l_{0}=%
\sqrt{\hbar /m\omega }$ is the oscillator length and $N$ is the number of
particle. Correspondingly, in the regime of a Boltzmann gas, the requirement
reads: $(\hbar \omega /T)(d_{\mathrm{phase}}/\lambda _{T})\gg 1$, or $d_{%
\mathrm{phase}}/l_{0}\gg \sqrt{T/\hbar \omega }$.                                             
          
\begin{figure}[tbp]
\resizebox{2.9in}{5.0in}
{\includegraphics{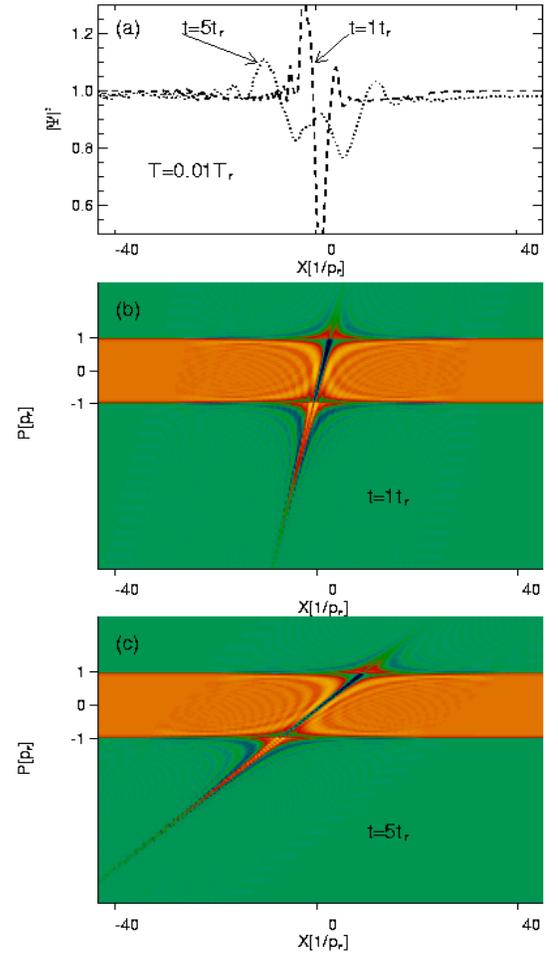}}
\caption{The time evolution of the gas density and the Wigner function after the
phase imprinting with $d_{\mathrm{phase}}=2\lambda _{F}$: (a) the gas
density at the times $t=t_{F}$ (the dashed curve) $t=5t_{F}$ (the dotted
curve); (b) the Wigner function at the time $t=t_{F}$; (c) the Wigner
function at the time $t=5t_{F}$. The temperature of the gas $T=0.01T_{F}$.}
\label{fig3}
\end{figure}

\begin{figure}[tbp]
\resizebox{2.9in}{5.0in}
{\includegraphics{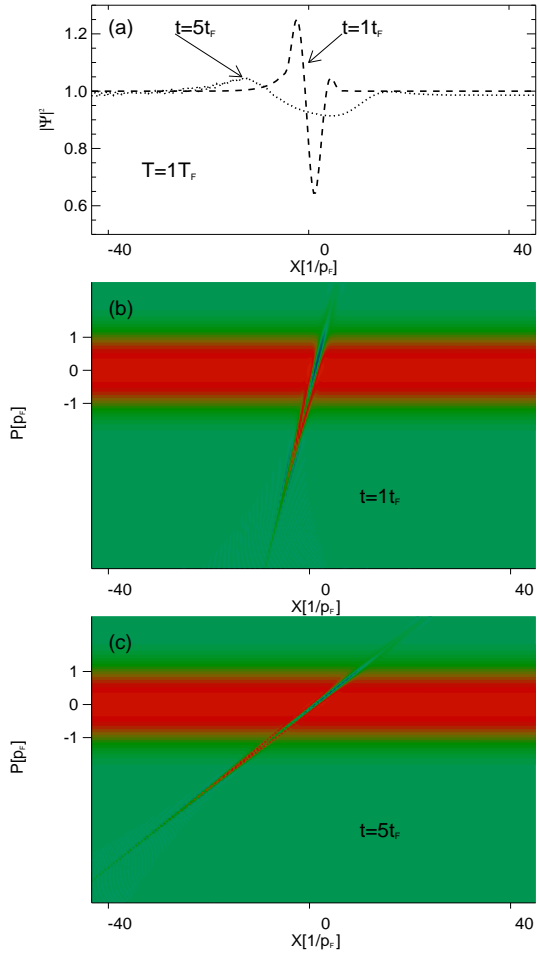}}
\caption{The same as in Fig. \ref{fig1} but for the temperature
$T=T_{F\text{.}}$} 
\label{fig4}
\end{figure}

In the regime where $d_{\mathrm{phase}}\lesssim \lambda _{F},\ \lambda _{T}$%
, the excitations created during the phase imprinting have a very broad
distribution with the width $\delta p\sim \hbar /d_{\mathrm{phase}}\gtrsim
p_{F},\ p_{T}$ in a spatially homogeneous gas, and $\delta \varepsilon \sim
\hbar \omega (l_{0}/d_{\mathrm{phase}})^{2}$ in a trapped one. In this case
one would expect a very rapid decay of ''solitons'' on a time scale $\tau
\sim [\hbar /\max (T_{F},T)][d_{\mathrm{phase}}/\min (\lambda _{F},\lambda
_{T})]^{2}\ll \hbar /T_{F}$ for both spatially homogeneous and trapped Fermi
gases. The corresponding numerical results for the density and Wigner 
function evolution after the $2\pi$--step phase imprinting with 
$d_{\mathrm{phase}}=2 \lambda_F$ are shown in Figs. \ref{fig3} (for the 
temperature $T=0.01T_F$) and \ref{fig4} (for the temperature $T=T_F$). In the
quantum  degenerate case (Fig. \ref{fig3}), the Wigner function has a well
pronounced  fringes structure, and the corresponding complicated structure in 
the density profile decays rather fast. For the non--degenerate gas (Fig.
\ref{fig4}),  the Wigner function is smooth everywhere except for the narrow 
region subjected by the phase imprinting. In this case, the density profile 
has no small-scale structures, as compared to the degenerate case from Fig. 
\ref{fig3}, but also relaxes very fast to the uniform density distribution.

\section{Density matrix formalism}

\label{DenMatrix}

Another way to discuss the soliton's formation in a gas of neutral fermionic
atoms is to start from a set of equations for reduced density matrices. In
particular, the equation of motion for the one--particle density matrix
involves the two--particle density matrix $\rho_2$ and is given by \cite{equ}
\begin{eqnarray}
i\hbar \frac{\partial}{\partial t} \rho_1(\vec{r}_1,\vec{r}_2,t) = -\frac{%
\hbar^2}{2 m} (\vec{\nabla}_1^2 - \vec{\nabla}_2^2) \, \rho_1(\vec{r}_1,\vec{%
r}_2,t) \phantom{11111} & &  \nonumber \\
+ \int d^3 r^{\prime} \left[V(\vec{r}_1-\vec{r}^{\,\prime}) - V(\vec{r}_2-%
\vec{r}^{\,\prime})\right] \rho_2(\vec{r}_1,\vec{r}^{\,\prime};\vec{r}_2,%
\vec{r}^{\,\prime},t) & &  \nonumber \\
+ \left[V_{ext}(\vec{r}_1,t)-V_{ext}(\vec{r}_2,t)\right] \rho_1(\vec{r}_1,%
\vec{r}_2,t) \;, \phantom{11111111111l} & &  \label{densitymat}
\end{eqnarray}
where $V(\vec{r}_1-\vec{r}_2)$ is the two--particle interaction term and $%
V_{ext}(\vec{r},t)$ is the external potential. In the limit $\vec{r}_1
\rightarrow \vec{r}_2$ Eq. (\ref{densitymat}) leads to the continuity
equation 
\begin{eqnarray}
\frac{\partial \rho(\vec{r},t)}{\partial t} + \vec{\nabla} \cdot \left[\rho(%
\vec{r},t) \, \vec{v}(\vec{r},t)\right] & = & 0 \;,  \label{hydr1}
\end{eqnarray}
where the density and velocity fields are defined as follows: 
\begin{eqnarray}
\rho(\vec{r},t) & = & \lim_{\vec{r}_1 \rightarrow \vec{r}_2} \rho_1(\vec{r}%
_1,\vec{r}_2,t)  \nonumber \\
\vec{v}(\vec{r},t) & = & \frac{\hbar}{2 m} \lim_{\vec{r}_1 \rightarrow \vec{r%
}_2} (\vec{\nabla}_1 - \vec{\nabla}_2) \, \chi_1(\vec{r}_1,\vec{r}_2,t)
\label{deffield}
\end{eqnarray}
and $\chi_1(\vec{r}_1,\vec{r}_2,t)$ is the phase of the one--particle
density matrix.

One can also rewrite Eq. (\ref{densitymat}) introducing the center--of--mass
($\vec{R}=(\vec{r}_1+\vec{r}_2)/2$) and the relative position ($\vec{s}=\vec{%
r}_1-\vec{r}_2$) coordinates. By taking the derivative of Eq. (\ref
{densitymat}) with respect to the coordinate $\vec{s}$ the hydrodynamic
Euler--type equation of motion is obtained in the limit $\vec{s} \rightarrow
0$ 
\begin{eqnarray}
\frac{\partial \vec{v}(\vec{r},t)}{\partial t} & = & -\frac{\vec{\nabla}
\cdot \mathrm{T}}{m\, \rho(\vec{r},t)} - [\vec{v}(\vec{r},t) \cdot \vec{%
\nabla}] \, \vec{v}(\vec{r},t)  \nonumber \\
& + & \frac{\vec{F}_{int}(\vec{r},t)}{m\, \rho(\vec{r},t)} -\frac{\vec{\nabla%
} V_{ext}(\vec{r},t)}{m} \;.  \label{hydr2}
\end{eqnarray}
Here, the kinetic--energy stress tensor $\mathrm{T}$, whose elements are
given by 
\begin{eqnarray}
\mathrm{T}_{kl} & = & -\frac{\hbar^2}{m} \lim_{\vec{s} \rightarrow 0} \frac{%
\partial^{\,2} \sigma_1(\vec{R},\vec{s},t)} {\partial s_k \partial s_l} \;,
\label{kin}
\end{eqnarray}
depends on $\sigma_1(\vec{r}_1,\vec{r}_2,t)$, the amplitude of one--particle
density matrix and can be calculated based on a local equilibrium assumption
which is the substance of Thomas--Fermi approximation. $\vec{F}_{int}(\vec{r}%
,t)$, on the other hand, represents the force due to fermion--fermion
interaction (for spin--polarized atomic fermions the dipole--dipole
interaction is a good candidate) and is defined as 
\begin{eqnarray*}
\vec{F}_{int}(\vec{r},t) & = & -\int d^3 r^{\prime} \, \vec{\nabla}_{\vec{r}%
} \, V(\vec{r}-\vec{r}^{\,\prime}) \, \rho_2(\vec{r},\vec{r}^{\,\prime};\vec{%
r},\vec{r}^{\,\prime},t) \;.
\end{eqnarray*}

The one--particle Wigner function within the Thomas--Fermi approximation is
given by 
\begin{eqnarray*}
w(\vec{r}, \vec{p}) & = & \eta(\hbar^2[6\pi^2 \rho(\vec{r})] ^{2/3} - \vec{p}%
^{\; 2}) \;,
\end{eqnarray*}
where $\eta()$ is Heaviside's unit step function and hence the one--particle
density matrix is calculated as 
\begin{eqnarray*}
\rho_1(\vec{R}, \vec{s}) & = & \int \frac{d^3 p}{(2\pi \hbar)^3} w(\vec{R}, 
\vec{p}) e^{i \vec{p} \vec{s} / \hbar} \;.
\end{eqnarray*}
After straightforward algebra one obtains: 
\begin{eqnarray*}
\rho_1(\vec{R}, \vec{s}) & = & \frac{2}{(2\pi \hbar)^2} \frac{1}{s} \left[ - 
\frac{\hbar p_F}{s} \cos \left(\frac{p_F s}{\hbar} \right) + \frac{\hbar^2}{%
s^2} \sin \left(\frac{p_F s}{\hbar} \right) \right] \;.
\end{eqnarray*}
The one--particle density matrix is a real function in this case and depends
only on the length $s=|\vec{s}|$ of vector $\vec{s}$. In the limit $\vec{s}
\rightarrow 0$ the kinetic--energy stress tensor $\mathrm{T}$ is getting
diagonal with $(\hbar^2/m) (1/30/\pi^2) (6 \pi^2 \rho)^{5/3}$ term at each
position. Then, assuming the case of non--interacting fermions
(spin--polarized atoms at low temperature), the Eqs. (\ref{hydr1}) and (\ref
{hydr2}) become a closed set of equations for the density and velocity
fields: 
\begin{eqnarray}  \label{hydro}
\frac{\partial \rho}{\partial t} + \vec{\nabla} \cdot \left( \rho \vec{v}
\right) = 0 \phantom{aaaaaaaaaaaaaaaaaaaaaaaaaaaaa} & &  \nonumber \\
\frac{\partial \vec{v}}{\partial t} + (\vec{v} \cdot \vec{\nabla}) \, \vec{v}
+ \vec{\nabla} \left( \frac{\hbar^2}{2 m^2} (6 \pi^2)^{3/2}\; \rho^{2/3} + 
\frac{V_{\mathrm{ext}}}{m} \right) = 0 \;, & &  \nonumber \\
\end{eqnarray}

Note that the Eqs. (\ref{hydr1}) and (\ref{hydr2}) are general and could be
used also for a bosonic system. For example, for the system of bosons
described by the many--body wave function $\Psi(\vec{r}_1,...,\vec{r}%
_N)=\varphi(\vec{r}_1)\cdot...\cdot \varphi(\vec{r}_N)$ (as in the case of
the Bose--Einstein condensate in the mean--field approximation) the
one--particle density matrix is given by 
\begin{eqnarray*}
\rho_1(\vec{R}, \vec{s}) & = & \varphi(\vec{R}+\frac{1}{2} \vec{s}) \;
\varphi^*(\vec{R}-\frac{1}{2} \vec{s}) \;.
\end{eqnarray*}
Now, the kinetic--energy stress tensor possesses off--diagonal elements and
leads to the so--called ``quantum pressure'' term in the equation of motion.
Assuming contact interaction between the bosonic atoms dominates, the
appropriate equations are written as 
\begin{eqnarray}
\frac{\partial \rho}{\partial t} + \vec{\nabla} \cdot \left( \rho \vec{v}
\right) = 0 \phantom{aaaaaaaaaaaaaaaaaaaaaaaaaaaaaa} & &  \nonumber \\
\frac{\partial \vec{v}}{\partial t} + \vec{\nabla} \left( \frac{4 \pi
\hbar^{2} a}{m^{2}} \rho + \frac{\vec{v}^{\:2}}{2} + \frac{V_{\mathrm{ext}}}{%
m} - \frac{\hbar^{2}}{2 m^{2}} \frac{\vec{\nabla}^{2} \sqrt{\rho}}{\sqrt{\rho%
}} \right) = 0 \phantom{aa}  \label{GPmadelung}
\end{eqnarray}
and are the hydrodynamic \cite{Madelung} representation of the
Gross--Pitaevskii equation.

It is easy to check that for the spatially homogeneous system, the
one--dimensional counterpart of Eqs. (\ref{hydro}) has the following
solution 
\begin{eqnarray}  \label{TFsolution}
\rho(x,t) & = & \rho_{0} + \frac{1}{2 \pi} [\phi ^{\prime
}(x+(p_{F}/m)t)-\phi ^{\prime} (x-(p_{F}/m)t)]  \nonumber \\
v(x,t) & = & - \frac{\hbar}{2 m} [\phi ^{\prime }(x+(p_{F}/m)t) + \phi
^{\prime}(x-(p_{F}/m)t)]  \nonumber \\
\end{eqnarray}
assuming the slope of the phase being imprinted is much less than the
unperturbed density ($\phi ^{\prime} \ll \rho_{0}$). This assumption is
equivalent to the condition $d_{\mathrm{phase}}\gg \lambda _{F}$ considered
by us in Sec. \ref{Wigner} and the solution (\ref{TFsolution}) coincides
with expression (\ref{density3}). It also coincides with the Thomas--Fermi
approximation, valid in one--dimensional space when the number of atoms is
large enough (Sec. \ref{Schrodinger}). Again, the solution (\ref{TFsolution})
represents two quasisolitons (a bright and a dark one) traveling in opposite
directions after imprinting (at $t=0$) the phase $\phi(x)$. The speed of 
quasisolitons is equal to the speed of sound. A similar solution can be found 
also in two-- and three--dimensional case 
\begin{eqnarray}
\rho(t) & = & \rho_{0} + \alpha [\phi ^{\prime }(x + c t)-\phi ^{\prime} (x
- c t)]  \nonumber \\
v_x(t) & = & - \frac{\hbar}{2 m} [\phi ^{\prime }(x + c t) + \phi
^{\prime}(x - c t)] \;,  \label{TFsol2D3D}
\end{eqnarray}
where $c$ equals the speed of sound, $\alpha \sim \rho_{0}^{1/2}$ or $\alpha
\sim \rho_{0}^{2/3}$ respectively in two-- and three--dimensional space, and
the phase being imprinted changes only in '$x$' direction. Again, the
solutions (\ref{TFsol2D3D}) are valid in the limit $d_{\mathrm{phase}}\gg
\lambda _{F}$.

It is clear from Fig. \ref{1Ddecay} that in the regime where $d_{\mathrm{%
phase}}\lesssim \lambda _F$, the dynamics of fermionic system should be
discussed rather in terms of soliton--like structures with dispersion
leading to the fast broadening of pulses. Even at zero temperature the 
phase imprinting excites states with energy above Fermi level, depending 
on the width of the phase step $d_{\mathrm{phase}}$.The deviation of the 
group velocity of bright and dark wave packets from the zero--temperature
sound velocity $p_F/m$ maybe qualitatively understood by the following
argument. In the spirit of the local density approximation one considers
local value of the Fermi momentum determined by the local value of the
density. Since the local density is lower at the center of dark quasisoliton
its velocity is thus reduced. Of course, the opposite is true for the bright
quasisoliton. Using Eq. (\ref{TFsolution}) one gets local corrections to
$p_F$ of the form $\delta p_F = \mp \frac{\hbar}{2} \phi ^{\prime } (0)$. This
correction agrees with numerical results in the order of magnitude. More
precise estimation of the asymmetry of velocities requires, however,
taking into account dispersion effects and systematic expansion in
$\lambda_F/d_{\mathrm{phase}}$ up to higher orders.

There is an equivalent way of calculating the kinetic--energy stress tensor $%
\mathrm{T}$ (\ref{kin}), involving the one--particle Wigner function 
\begin{eqnarray}
\mathrm{T}_{kl} & = & \int \frac{p_k p_l}{m} w(\vec{r}, \vec{p})\, d^3 p \,,
\label{kinWigner}
\end{eqnarray}
which is, however, more suitable for the finite--temperature extension of
the Thomas--Fermi approach. In equilibrium, the Wigner function for a
uniform system is just the Fermi--Dirac distribution function $%
f(\varepsilon)=(\exp(\beta(\varepsilon - \mu))+1)^{-1}$ with $%
\varepsilon=p^2/2m$ and $\mu$ being the particle's kinetic energy and the
chemical potential (constant for a homogeneous case) respectively. The
Thomas--Fermi approximation assumes, as usually, that the system is locally
uniform and the appropriate equations (in one--dimensional form) are written
as 
\begin{eqnarray}  \label{TFtemp}
\frac{\partial \rho}{\partial t} + \frac{\partial}{\partial x} \left( \rho v
\right) = 0 \phantom{aaaaaaaaaaaaaaaaaaaaaaaaaaaaaa} & &  \nonumber \\
\frac{\partial v}{\partial t} + v \frac{\partial v}{\partial x} + \frac{1}{m
\rho} \frac{\partial}{\partial x} \frac{1}{2\pi\hbar} \int\limits_{-%
\infty}^{\infty} \frac{p^2 / m \, dp}{\exp(\beta(\frac{p^2}{2m} - \mu))+1} =
0 \;. & &  \nonumber \\
\end{eqnarray}
Here, the chemical potential is position--dependent and is calculated from
the relation 
\begin{eqnarray}
\rho=\frac{1}{2\pi\hbar} \int\limits_{-\infty}^{\infty} \frac{d p}{%
\exp(\beta(\frac{p^2}{2m} - \mu))+1} \;.  \label{chempot}
\end{eqnarray}

We consider now the case when the temperature is finite but the gas is still
strongly degenerate, i.e. $T \ll T_F$. By using the Sommerfeld expansion and
keeping only the first term one can easily find the expression
$\mu=(2\pi\hbar)^2/(8m)\,\rho^2$. It turns out that in the regime where the
imprinted phase changes slowly in comparison with unperturbed density, the
finite--temperature time--dependent Thomas--Fermi model defined by the Eqs. (%
\ref{TFtemp}) and (\ref{chempot}) has the following solution 
\begin{eqnarray}  \label{TFtempsol}
\rho = \rho_{0} + \int\limits_0^{\infty} \frac{d \varepsilon}{2 \pi} \left( -%
\frac{\partial f}{\partial \varepsilon} \right) [\phi ^{\prime }(x+\sqrt{%
\frac{2\varepsilon}{m}}t)- \phi ^{\prime} (x-\sqrt{\frac{2\varepsilon}{m}}%
t)] & &  \nonumber \\
v = - \frac{\hbar}{2 m} \int\limits_0^{\infty} d \varepsilon \left( -\frac{%
\partial f}{\partial \varepsilon} \right) [\phi ^{\prime }(x+\sqrt{\frac{%
2\varepsilon}{m}}t) + \phi ^{\prime}(x-\sqrt{\frac{2\varepsilon}{m}}t)] & & 
\nonumber \\
\end{eqnarray}

For the temperature $T=0$ one has $(-\partial f / \partial
\varepsilon)=\delta(\varepsilon- \varepsilon_F)$ and the solution (\ref
{TFtempsol}) reduces to the one found previously (\ref{density3}, \ref{TFsolution}). 
To verify the ansatz (\ref{TFtempsol}) we put it in the set of Eqs. (\ref
{TFtemp}) and keep only linear terms in $\phi^{\prime}$ (according to $d_{%
\mathrm{phase}}\gg \lambda _{F}$). The solvability condition has then the
form 
\begin{eqnarray}
\int\limits_0^{\infty} d \varepsilon \; \left( -\frac{\partial f}{\partial
\varepsilon} \right) \left[ \sqrt{\frac{2\varepsilon}{m}}-\sqrt{\frac{%
2\varepsilon_F}{m}}\, \right] \times \phantom{aaaaaa} & &  \nonumber \\
\left[ \phi^{\prime\prime}\left(x+\sqrt{\frac{2\varepsilon}{m}}t\right) \pm
\phi^{\prime\prime}\left(x-\sqrt{\frac{2\varepsilon}{m}}t\right) \right] =0
& &  \label{solvability}
\end{eqnarray}
and is true for low temperatures $T\ll T_F$ since then the derivative of the
Fermi--Dirac distribution function strongly peaks at $\varepsilon=%
\varepsilon_F$, whereas the rest of the integrand is very small. The
solution (\ref{TFtempsol}) is identical with (\ref{density4}).

\begin{table}[tbp]
\caption{Speeds of dark and bright quasisolitons (in units of speed of sound)
generated in a one--dimensional homogeneous Fermi gas of various number of
atoms after imprinting a single phase step of $\pi$ and the width of $2.5\,
\lambda_F$}
\label{table2}%
\begin{ruledtabular}
\begin{tabular}{rcc}
N & $\frac{v_b}{c}$ & $\frac{v_d}{c}$ \\
\hline
500   & -1.061 & +0.937 \\
1000  & -1.038 & +0.965 \\
5000  & -1.010 & +0.993 \\
8000  & -1.006 & +0.994 \\
\end{tabular}
\end{ruledtabular}
\end{table}

\subsection{Numerical results}

We have solved numerically the set of Eqs. (\ref{hydro}). To this end, we
applied the inverse Madelung transformation \cite{inverse} to Eqs. (\ref
{hydro}) and used the Split-Operator technique. Imaginary--time propagation
method was chosen to generate the ground state of a nonuniform (trapped)
Fermi gas. Real--time propagation, on the other hand, allows to investigate
the phase imprinting and developing instabilities after that.

First, by imprinting the phase on a one--dimensional uniform system we again
investigate the regime where $d_{\mathrm{phase}}\gg \lambda _{F}$, however,
this time by increasing the number of atoms while keeping the same the width
of the phase step. Of course, larger number of atoms means smaller Fermi
length. Increasing the number of atoms is easily attainable within
Thomas--Fermi approach as opposed to the direct solution of the many--body
Schr\"odinger equation. Results regarding the speeds of bright and dark
quasisolitons are presented in Table \ref{table2}. It is clear that reaching the
above mentioned regime forces both quasisolitons to move with the same velocity
equal to the speed of sound.

\section{Conclusions}
\label{conclusions}
In this paper we have studied in detail the possibility of generation of
soliton--like structures in ultracold Fermi gases using the method of
phase imprinting. Generation of such structures is feasible with existing
experimental setups. The soliton--like structures are generated because in 
the process of phase imprinting in the limit in which the characteristic
length of the imprint $d_{\mathrm{phase}} \ll \lambda_F$ only a narrow band 
of states with momenta close to $\pm p_F$ are excited. The life time of 
soliton--like structures can be quite long even at not extremely low 
temperatures allowing for observations. We have presented both numerical 
results and analytic theory explaining the mechanism of generation and 
evolution of soliton--like structures both in homogeneous quasi 1D cylinders 
and in quasi 1D harmonic traps. It will be interesting to generalize the 
presented theory to the case of two--component Fermi gas in normal phase.

\acknowledgments
We thank C. Salomon, J. Dalibard, L. Santos, A. Sanpera.
This paper has been supported by Deutsche Forschungsgemeinschaft 
(among others in the frame of SFB 407, SP 1116, GK 282, and 436POL),
TRN 'Cold Atoms and Molecules', ESF PESC Programm 'BEC 2000+'.
K.R. acknowledges support of the Subsidy by Foundation for Polish Science.
Part of the results has been obtained using computers at the
Interdisciplinary Center for Mathematical and Computational Modeling at
Warsaw University.

\end{document}